\documentclass[letter]{IEEEtran}
\usepackage{cmap}
\usepackage{amsmath}
\usepackage{amsfonts}
\usepackage{subfigure}
\usepackage{graphics,graphicx,picture}
\usepackage{algorithm}
\usepackage{algorithmic}
\usepackage{mathrsfs}
\usepackage{cases}
\usepackage{bm, comment,epstopdf,color,amssymb}
\usepackage{cite}

\begin{document}

\title{Secure Beamforming in MISO NOMA Backscatter Device Aided Symbiotic Radio Networks}

\author{Yiqing Li, Miao Jiang, Qi Zhang, \emph{Member}, \emph{IEEE}, and Jiayin Qin
	
\thanks{}

}

\markboth{}
{Li \MakeLowercase{\textit{et al.}}: Secure Beamforming in MISO NOMA Backscatter Device Aided Symbiotic Radio Networks}

\maketitle
\begin{abstract}
Symbiotic radio (SR) networks are possible solutions to the future low-power wireless communications for massive Internet of Things devices. In this paper, we investigate a multiple-input-single-output non-orthogonal multiple access (NOMA) backscatter device (BD) aided SR network with a potential eavesdropper. In the network, a base station (BS) broadcasts signals to a central user and a cell-edge user using the NOMA protocol. With ambient backscatter modulation, the BD transmits its own messages to the central user over incident signals from the BS. We propose a constrained concave convex procedure-based algorithm which maximizes the $\epsilon$-outage secrecy rate from the BD to the central user under the achievable secrecy rate constraints from the BS to the central and cell-edge users. Simulation results illustrate that our proposed network achieves a much larger secrecy rate region than the orthogonal multiple access (OMA) network.
\end{abstract}
\IEEEpeerreviewmaketitle

\section{introduction}
Compared with conventional communication technology, ambient backscatter communication (AmBC) has high spectrum and energy efficiency to support future low-power Internet of Things (IoT) devices \cite{VLiu,DTHoang}. In AmBC, the passive IoT device called backscatter device (BD) transmits its own messages by modulating them over its received ambient radio frequency (RF) signals. To allow the backscattered signals share the spectrum of primary system, symbiotic radio (SR) networks were proposed in \cite{RLong,RLong1,QZhang,HGuo,HGuo1}.

In \cite{QZhang}, a non-orthogonal multiple access (NOMA) downlink BD aided SR network was investigated. In the network, a base station (BS) transmits signals to a central user and a cell-edge user using the NOMA protocol, while a BD transmits its own messages to the central user by modulating them over its received ambient RF signals.

Because of the openness of the wireless transmission medium, wireless information is susceptible to eavesdropping \cite{Yzhang,YLiu,YLi,MJiang}. In this paper, we consider that there exists a potential eavesdropper in the multiple-input-single-output (MISO) NOMA downlink BD aided SR network. The potential eavesdropper eavesdrops the signals transmitted from the BS as well as those from the BD. The scenario is typical for wireless communications in an unmanned supermarket, such as Amazon Go. In the unmanned supermarket, a wireless access point (AP) supports the wireless communications for a self-checkout machine (an entrusted central user), a sweeper robot (an entrusted cell-edge user), a grocery item with a radio frequency identification (RFID) tag (an entrusted BD), and an anonymous user (a potential eavesdropper).

In this paper, our objective is to maximize the $\epsilon$-outage secrecy rate  \cite{Tse} from the BD to the central user under the achievable secrecy rate constraints from the BS to the central and cell-edge users. The formulated problem contains an $\epsilon$-outage probability constraint. We propose to equivalently transform the $\epsilon$-outage probability constraint into two constraints. We employ the eigenvalue decomposition (EVD) to recast aforementioned constraints into difference-of-convex (DC) constraints. Using rank-one relaxation and constrained concave convex procedure (CCCP), we propose to solve a sequence of convex semidefinite programmings (SDPs) to obtain the solution to the original problem.

\emph{Notations}: Boldface lowercase and uppercase letters denote vectors and matrices, respectively. The conjugate transpose, trace, and determinant of the matrix $\mathbf{A}$ are denoted as $\mathbf{A}^{H}$, $\mbox{tr}(\mathbf{A})$, and $\mbox{det}(\mathbf{A})$, respectively. By $\mathbf{A}\succeq\mathbf{0}$, we mean that $\mathbf{A}$ is positive semidefinite. $\mathcal{CN}(0,\sigma^2)$ denotes the distribution of a circularly symmetric complex Gaussian random variable with zero mean and variance $\sigma^2$. $\|\mathbf{a}\|^2$ denotes the squared Euclidean norm of a complex vector $\mathbf{a}$.

\section{System Model and Problem Formulation}

\begin{figure}
\centering
\includegraphics[width=3.4in]{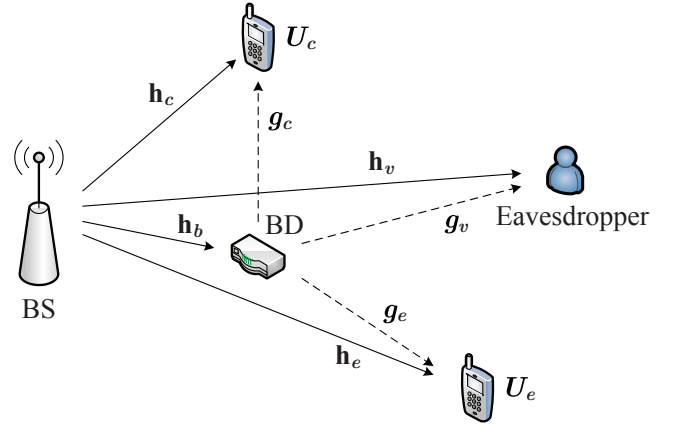}
\caption{System model of an MISO NOMA downlink BD aided SR network with a potential eavesdropper.}
\end{figure}

As shown in Fig. 1, consider an MISO NOMA downlink BD aided SR network with a potential eavesdropper, which consists of five nodes, namely, one BS, one BD, one central user denoted as $U_c$, one cell-edge user denoted as $U_e$, and a potential eavesdropper. The BS is equipped with $M\geq1$ antennas. Each of the BD, the central user $U_c$, the cell-edge user $U_e$, and the eavesdropper is equipped with a single antenna. The BS employs the NOMA protocol to transmit signals to $U_c$ and $U_e$ simultaneously, which are expressed as
\begin{align}
\mathbf{x}= \mathbf{w}_cs_c + \mathbf{w}_es_e
\end{align}
where $s_c\sim \mathcal{CN}\left(0, 1\right)$ and $s_e\sim \mathcal{CN}\left(0, 1\right)$ denote the signals intended to $U_c$ and $U_e$, respectively; $\mathbf{w}_c\in \mathbb{C}^{M \times 1}$ and $\mathbf{w}_e\in \mathbb{C}^{M \times 1}$ denote the beamforming vectors for $U_c$ and $U_e$, respectively.

The BD, after receiving the incident signals from the BS, modulates them with its own symbol $s_b$, where $\mathbb{E}[\left|s_b\right|^2] = 1$, and backscatters the modulated signals. The symbol $s_b$ is intended for $U_c$. Thus, the received signal at $U_i$, $i\in\{c,e\}$ is
\begin{align} \label{eq3}
y_i=\mathbf{h}_i^H\mathbf{x}+ \sqrt{\alpha}g_is_b\mathbf{h}_b^H\mathbf{x}+ n_i
\end{align}
where $\mathbf{h}_c\in \mathbb{C}^{M \times 1}$, $\mathbf{h}_e\in \mathbb{C}^{M \times 1}$, and $\mathbf{h}_b\in \mathbb{C}^{M \times 1}$ denote the block flat-fading channels from the BS to $U_c$, $U_e$, and the BD, respectively; $g_c\in \mathbb{C}^{1 \times 1}$ and $g_e\in \mathbb{C}^{1 \times 1}$ denote the block flat-fading channels from the BD to $U_c$ and $U_e$, respectively; $\alpha\in[0,1]$ denotes the reflection coefficient; $n_c\sim \mathcal{CN}\left(0, \sigma^2\right)$ and $n_e\sim \mathcal{CN}\left(0, \sigma^2\right)$ denote the additive Gaussian noises at $U_c$ and $U_e$, respectively.

Affected by the double fading effect \cite{QZhang}, in \eqref{eq3}, the backscatter link signal from the BD, i.e., $\sqrt{\alpha}g_is_b\mathbf{h}_b^H\mathbf{x}$,
is much weaker than the direct link from the BS, i.e., $\mathbf{h}_i^H\mathbf{x}$. To ensure that the backscatter link signal can be correctly decoded, $U_c$ first decodes $s_e$, then decodes $s_c$, and finally decodes $s_b$, using successive interference cancellation (SIC) technique. Therefore, the signal-to-interference-plus-noise ratio (SINR) to decode $s_e$ is
\begin{align}
\gamma_{c,e} = \frac{\left|\mathbf{h}_c^H \mathbf{w}_e \right|^2}{\sigma^2 + \left|\mathbf{h}_c^H\mathbf{w}_c \right|^2 + \alpha f \left|g_c\right|^2}
\end{align}
where $f=|\mathbf{h}_b^H\mathbf{w}_c|^2+|\mathbf{h}_b^H\mathbf{w}_e|^2$. If $s_e$ is decoded successfully, after SIC, the SINR to decode $s_c$ is
$\gamma_{c,c}=\left|\mathbf{h}_c^H \mathbf{w}_c\right|^2/(\sigma^2 + \alpha f\left|g_c\right|^2)$. If both $s_e$ and $s_c$ are decoded successfully, after SIC, the SINR to decode $s_b$ is
\begin{equation}
\gamma_{c,b} = \alpha \left|g_c\right|^2\left(\left|\mathbf{h}_b^H\mathbf{w}_cs_c+\mathbf{h}_b^H\mathbf{w}_es_e\right|^2\right)/\sigma^2.
\end{equation}

At the potential eavesdropper, the SINR to decode $s_c$ is
\begin{align}
\gamma_{v,c} = \frac{\left|\mathbf{h}_v^H \mathbf{w}_c\right|^2}{\sigma^2 + \left|\mathbf{h}_v^H\mathbf{w}_e\right|^2 + \alpha f\left|g_v\right|^2}
\end{align}
where $\mathbf{h}_v\in \mathbb{C}^{M \times 1}$ and $g_v\in \mathbb{C}^{1 \times 1}$ denote the block flat-fading channels from the BS and BD to the eavesdropper, respectively. Therefore, the achievable secrecy rate for signals from the BS to $U_c$ is
\begin{align}
R_c=\log_2(1+\gamma_{c,c})-\log_2(1+\gamma_{v,c}).
\end{align}
Similarly, the achievable secrecy rate for signals from the BD to $U_c$ is
$R_b=\log_2(1+\gamma_{c,b})-\log_2(1+\gamma_{v,b})$ where
\begin{align}
\gamma_{v,b} = \frac{\alpha f\left|g_v\right|^2}{\sigma^2 + \left|\mathbf{h}_v^H \mathbf{w}_c\right|^2 + \left|\mathbf{h}_v^H\mathbf{w}_e\right|^2}.
\end{align}

At the cell-edge user $U_e$, the SINR to decode $s_e$ is
\begin{align}
\gamma_{e,e} = \frac{\left|\mathbf{h}_e^H\mathbf{w}_e\right|^2}{\sigma^2 + \left|\mathbf{h}_e^H\mathbf{w}_c\right|^2 + \alpha f\left|g_e\right|^2}.
\end{align}
Thus, the achievable secrecy rate for signals from the BS to $U_e$ is $R_e=\log_2(1+\gamma_{e,e})-\log_2(1+\gamma_{v,e})$ where
\begin{align}
\gamma_{v,e} = \frac{\left|\mathbf{h}_v^H \mathbf{w}_e\right|^2}{\sigma^2 + \left|\mathbf{h}_v^H\mathbf{w}_c\right|^2 + \alpha f \left|g_v\right|^2}.
\end{align}

Our goal is to maximize the $\epsilon$-outage secrecy rate \cite{Tse} from the BD to $U_c$ under the achievable secrecy rate constraints from the BS to $U_c$ and $U_e$, which is formulated as
\begin{subequations}\label{bq1}
\begin{align}
\label{bq1a}\max_{\mathbf{w}_c, \mathbf{w}_e, r_b\geq0} & r_b \\
\label{bq1b} \mbox{s.t.}\quad\ & R_c\geq r_c,\ R_e\geq r_e,\ R_{c,e}\geq r_e,\\
\label{bq1c} & \mbox{Pr} \left(R_b \geq r_b\right)\geq 1 - \epsilon,\\
\label{bq1d} & \left\|\mathbf{w}_1 \right\|^2 + \left\|\mathbf{w}_2\right\|^2 \leq P
\end{align}
\end{subequations}
where $R_{c,e}=\log_2(1+\gamma_{c,e})-\log_2(1+\gamma_{v,e})$; $r_c$ and $r_e$ denote the achievable secrecy rate constraints for signals from the BS to $U_c$ and $U_e$, respectively; $r_b$ denotes the $\epsilon$-outage secrecy rate for signals from the BD to $U_c$; $\epsilon$ denotes the target outage probability for signals from the BD to $U_c$; and $P$ denotes the transmit power constraint at the BS.

\section{Proposed CCCP-Based $\epsilon$-Outage Secrecy Rate Optimization Algorithm}
Introducing a slack variable $\zeta$, constraint \eqref{bq1c} can be rewritten as
\begin{align}\label{bq2}
\mbox{Pr} \left(\mathbf{s}^H \mathbf{Q} \mathbf{s} \geq \xi\right)&\geq 1 - \epsilon,\\
\label{bq3}\gamma_{v,b}&\leq \zeta,
\end{align}
where $\mathbf{s} = \left[s_c, s_e\right]^T$, $\xi= \sigma^2\left(\omega(\zeta+1)- 1\right)/(\alpha|g_c|^2)$, $\omega=2^{r_b}$, and
\begin{equation}
\mathbf{Q} = \begin{bmatrix}
\mathbf{h}_b^H\mathbf{w}_c\mathbf{w}_c^H \mathbf{h}_b & \mathbf{h}_b^H\mathbf{w}_c\mathbf{w}_e^H \mathbf{h}_b \\
\mathbf{h}_b^H\mathbf{w}_e\mathbf{w}_c^H \mathbf{h}_b &
\mathbf{h}_b^H\mathbf{w}_e\mathbf{w}_e^H \mathbf{h}_b
\end{bmatrix}.
\end{equation}
Constraint \eqref{bq2} is non-convex because of coupled optimization variables. Since $\mathbf{Q}$ is a Hermitian matrix and $\mbox{det}(\mathbf{Q})=0$, the eigenvalue decomposition (EVD) of matrix $\mathbf{Q}$ is $\mathbf{Q}=\mathbf{U}^H\mathbf{\Lambda}\mathbf{U}$, where $\mathbf{U}\in \mathbb{C}^{2 \times 2}$ is a unitary matrix and
\begin{equation}
\mathbf{\Lambda} = \begin{bmatrix}
\lambda & 0\\ 0 & 0
\end{bmatrix}
\end{equation}
in which $\lambda=\mathbf{h}_b^H\left(\mathbf{w}_c\mathbf{w}_c^H +  \mathbf{w}_e\mathbf{w}_e^H\right)\mathbf{h}_b$. Let $\bar{\mathbf{s}} = \mathbf{U}\mathbf{s}=\left[s_1, s_2\right]^T$. Since $\mathbf{s}\sim \mathcal{CN}\left(\mathbf{0}, \mathbf{I}\right)$, we have $\bar{\mathbf{s}}\sim \mathcal{CN}\left(\mathbf{0}, \mathbf{I}\right)$. Thus, constraint \eqref{bq2} is equivalent to
\begin{align}\label{bq5}
\mbox{Pr}\left(\lambda |s_1|^2 \geq \xi \right) \geq 1 - \epsilon.
\end{align}
Since $s_1\sim \mathcal{CN}\left(0, 1\right)$, $|s_1|^2$ is an exponential distributed random variable with parameter 1. We have
\begin{align}
\mbox{Pr}\left(\lambda |s_1|^2 \geq \xi \right)=\exp(-\xi/\lambda).
\end{align}
Constraint \eqref{bq5} is equivalent to $\lambda\geq  \xi/\rho$,
where $\rho=-\ln(1 - \epsilon)$, i.e.,
\begin{align}\label{bq8}
\mathbf{h}_b^H\left(\mathbf{w}_c\mathbf{w}_c^H +  \mathbf{w}_e\mathbf{w}_e^H\right)\mathbf{h}_b \geq \xi/\rho.
\end{align}
Let $\mathbf{W}_c=\mathbf{w}_c\mathbf{w}_c^H$ and $\mathbf{W}_e = \mathbf{w}_e\mathbf{w}_e^H$. Constraint \eqref{bq8} is equivalent to
\begin{align}\label{bq9}
\eta_1-\eta_2\leq\frac{\rho}{ \sigma^2} \mbox{tr}\left(\alpha|g_c|^2\mathbf{H}_b\left(\mathbf{W}_c + \mathbf{W}_e\right)\right)+1
\end{align}
where $\mathbf{H}_b=\mathbf{h}_b\mathbf{h}_b^H$, $\eta_1=\left(\omega + \zeta + 1\right)^2/2$, and $\eta_2=\omega^2/2+(\zeta + 1)^2/2$. Similarly, $\gamma_{v,b}\leq \zeta$ in \eqref{bq3} is equivalent to
\begin{align}\label{bq10}
\eta_3-\eta_4\geq\mbox{tr}\left(\alpha|g_v|^2\mathbf{H}_b\left(\mathbf{W}_c+\mathbf{W}_e\right)\right)
\end{align}
where
\begin{align}
\eta_3&=\left(\zeta + \sigma^2 + \mbox{tr}\left(\mathbf{H}_v\left(\mathbf{W}_c + \mathbf{W}_e\right)\right)\right)^2/2,\\
\eta_4&=\zeta^2/2+\left(\sigma^2 + \mbox{tr}\left(\mathbf{H}_v\left(\mathbf{W}_c + \mathbf{W}_e\right)\right)\right)^2/2.
\end{align}
Let $\mathbf{H}_c=\mathbf{h}_c\mathbf{h}_c^H$, $\mathbf{H}_e=\mathbf{h}_e\mathbf{h}_e^H$, and $\mathbf{H}_v=\mathbf{h}_v\mathbf{h}_v^H$. In \eqref{bq1b}, the expressions of $R_c$,  $R_e$, and $R_{c,e}$ can be transformed into
\begin{align}
R_c&=\left(\tau_1+\tau_2-\mu_1-\mu_2\right)/\ln2,\\
R_e&=\left(\tau_3+\tau_4-\mu_3-\mu_4\right)/\ln2,\\
R_{c,e}&=\left(\tau_5+\tau_6-\mu_5-\mu_6\right)/\ln2,
\end{align}
where
\begin{align}
\tau_j=\ln \left(\sigma^2 + \mbox{tr}\left(\mathbf{\Phi}_j\mathbf{W}_c + \mathbf{\Psi}_j\mathbf{W}_e\right) \right),\\
\mu_j=\ln \left(\sigma^2 + \mbox{tr}\left(\mathbf{\Sigma}_j\mathbf{W}_c + \mathbf{\Theta}_j\mathbf{W}_e\right) \right),
\end{align}
for $j\in\{1,2,\cdots,6\}$ in which
\begin{align}
&\mathbf{\Phi}_1=\mathbf{\Phi}_5=\mathbf{\Psi}_5=\mathbf{\Sigma}_5=\mathbf{H}_c+\alpha|g_c|^2\mathbf{H}_b,\\
&\mathbf{\Phi}_2=\mathbf{\Psi}_4=\mathbf{\Psi}_6=\alpha|g_v|^2\mathbf{H}_b, \quad \mathbf{\Theta}_3=\alpha|g_e|^2\mathbf{H}_b,\\
&\mathbf{\Phi}_3=\mathbf{\Psi}_3=\mathbf{\Sigma}_3=\mathbf{H}_e+\alpha|g_e|^2\mathbf{H}_b,\\ &\mathbf{\Phi}_4=\mathbf{\Phi}_6=\mathbf{\Psi}_2=\mathbf{\Sigma}_2=\mathbf{\Sigma}_4=\mathbf{H}_v+\alpha|g_v|^2\mathbf{H}_b,\\
&\mathbf{\Sigma}_6=\mathbf{\Theta}_2=\mathbf{\Theta}_4=\mathbf{\Theta}_6=\mathbf{H}_v+\alpha|g_v|^2\mathbf{H}_b,\\
&\mathbf{\Psi}_1=\mathbf{\Sigma}_1=\mathbf{\Theta}_1=\mathbf{\Theta}_5=\alpha|g_c|^2\mathbf{H}_b.
\end{align}
Problem \eqref{bq1} is equivalently recast as
\begin{subequations}\label{cq1}
\begin{align}
\label{cq1a}\max_{\mathbf{W}_c, \mathbf{W}_e, \omega\geq1, \zeta\geq0} & \omega \\
\label{cq1b} \mbox{s.t.}\quad\quad\ & \tau_1+\tau_2-\mu_1-\mu_2\geq r_c\ln2,\\
\label{cq1c}&\tau_3+\tau_4-\mu_3-\mu_4\geq r_e\ln2,\\
\label{cq1d}&\tau_5+\tau_6-\mu_5-\mu_6\geq r_e\ln2,\\
\label{cq1e}&\mathbf{W}_c\succeq\mathbf{0},\ \mathbf{W}_e\succeq\mathbf{0},\ \eqref{bq9},\ \eqref{bq10},\\
\label{cq1e}&\mbox{rank}(\mathbf{W}_c)=1,\ \mbox{rank}(\mathbf{W}_e)=1.
\end{align}
\end{subequations}
Employing the rank-one relaxation and omitting constraints \eqref{cq1e}, problem \eqref{cq1} is still non-convex because of the non-convex terms $\mu_j$, $j\in\{1,2,\cdots,6\}$, $\eta_2$, and $\eta_3$. We propose to replace the non-convex terms with their first-order Taylor expansions and solve a sequence of convex subproblems successively by employing the principle of CCCP.

Specifically, in the $(l + 1)$th iteration, given the optimal solution in the $l$th iteration, denoted as $\Omega^{(l)}=(\mathbf{W}_c^{(l)}, \mathbf{W}_e^{(l)}, \omega^{(l)}, \zeta^{(l)})$, the first-order partial derivatives of $\mu_j$, $j\in\{1,2,\cdots,6\}$, with respect to $\mathbf{W}_c$ and $\mathbf{W}_e$ are
\begin{align}
\frac{\partial\mu_j}{\partial\mathbf{W}_c}&= \frac{\mathbf{\Sigma}_j}{\sigma^2+\mbox{tr}\left(\mathbf{\Sigma}_j\mathbf{W}_c + \mathbf{\Theta}_j\mathbf{W}_e\right)},\\
\frac{\partial\mu_j}{\partial\mathbf{W}_e}&= \frac{\mathbf{\Theta}_j}{\sigma^2+\mbox{tr}\left(\mathbf{\Sigma}_j\mathbf{W}_c + \mathbf{\Theta}_j\mathbf{W}_e\right)}.
\end{align}
Thus, in the $(l + 1)$th iteration, $\mu_j$ can be replaced by its first-order Taylor expansion
\begin{align}
\mu_j&\leq\tilde{\mu}_j\left(\Omega^{(l)}\right)
=\ln \left(\sigma^2 + \mbox{tr}\left(\mathbf{\Sigma}_j\mathbf{W}_c^{(l)} + \mathbf{\Theta}_j\mathbf{W}_e^{(l)}\right) \right)\nonumber\\
&+\frac{\mbox{tr}\left(\mathbf{\Sigma}_j\left(\mathbf{W}_c-\mathbf{W}_c^{(l)}\right)+\mathbf{\Theta}_j\left(\mathbf{W}_e-\mathbf{W}_e^{(l)}\right)\right)}{\sigma^2+\mbox{tr}\left(\mathbf{\Sigma}_j\mathbf{W}_c^{(l)} + \mathbf{\Theta}_j\mathbf{W}_e^{(l)}\right)}.
\end{align}
Similarly, $\eta_2$ can be replaced by
\begin{align}
\eta_2&\geq \tilde{\eta}_2\left(\Omega^{(l)}\right)=\left(\omega^{(l)}\right)^2/2+ \left(\zeta^{\left(l\right)}+1 \right)^2/2\nonumber \\
&+\omega^{\left(l\right)}\left(\omega-\omega^{\left(l\right)}\right) + \left(\zeta^{\left(l\right)} + 1\right)\left(\zeta-\zeta^{\left(l\right)}\right)
\end{align}
and $\eta_3$ can be replaced by
\begin{align}
\eta_3&\leq \tilde{\eta}_3\left(\Omega^{(l)}\right)=\frac{1}{2} \left(\zeta^{\left(l\right)} + \sigma^2 +\varphi^{(l)}\right)^2\!+\!\left(\zeta^{\left(l\right)} + \sigma^2 +\varphi^{(l)}\right)\nonumber\\
&\cdot \left(\zeta - \zeta ^{\left(l\right)} + \mbox{tr}\left(\mathbf{H}_v \left(\mathbf{W}_c+ \mathbf{W}_e\right)\right)-\varphi^{(l)}\right)
\end{align}
where
$\varphi^{(l)}=\mbox{tr}(\mathbf{H}_v(\mathbf{W}_c^{(l)} + \mathbf{W}_e^{(l)}))$. Therefore, in the $(l + 1)$th iteration, we solve the following convex SDP
\begin{align}\label{cq10}
\max_{\mathbf{W}_c, \mathbf{W}_e, \omega\geq1, \zeta\geq0} & \omega \\
\mbox{s.t.}\quad\quad\ & \tau_1+\tau_2-\tilde{\mu}_1\left(\Omega^{(l)}\right)-\tilde{\mu}_2\left(\Omega^{(l)}\right)\geq r_c\ln2,\nonumber\\
&\tau_3+\tau_4-\tilde{\mu}_3\left(\Omega^{(l)}\right)-\tilde{\mu}_4\left(\Omega^{(l)}\right)\geq r_e\ln2,\nonumber\\
&\tau_5+\tau_6-\tilde{\mu}_5\left(\Omega^{(l)}\right)-\tilde{\mu}_6\left(\Omega^{(l)}\right)\geq r_e\ln2,\nonumber\\
&\mathbf{W}_c\succeq\mathbf{0},\ \mathbf{W}_e\succeq\mathbf{0},\ \eta_1-\tilde{\eta}_2\left(\Omega^{(l)}\right)\leq\phi,\nonumber\\
&\tilde{\eta}_3\left(\Omega^{(l)}\right)-\eta_4\geq\mbox{tr}\left(\alpha|g_v|^2\mathbf{H}_b\left(\mathbf{W}_c+\mathbf{W}_e\right)\right)\nonumber
\end{align}
where $\phi=(\rho/\sigma^2) \mbox{tr}\left(\alpha|g_c|^2\mathbf{H}_b\left(\mathbf{W}_c + \mathbf{W}_e\right)\right)+1$. After solving problem \eqref{cq10}, if the obtain solutions of $\mathbf{W}_c$ and $\mathbf{W}_e$ are not rank-one, the rank-one decomposition theorem proposed in \cite{WAi} can be applied to recover rank-one $\mathbf{W}_c$ and $\mathbf{W}_e$.

\section{Simulation Result}
In simulations, we assume that the BS is equipped with $M=4$ antennas. The channels are independent and identically distributed (i.i.d.) block flat Rayleigh fading channels such that $\mathbf{h}_c\sim\mathcal{CN}(\mathbf{0}, \mathbf{I})$, $\mathbf{h}_e\sim\mathcal{CN}(\mathbf{0},5^{-3}\mathbf{I})$,
$\mathbf{h}_b\sim\mathcal{CN}(\mathbf{0},\mathbf{I})$,
$\mathbf{h}_v\sim\mathcal{CN}(\mathbf{0},10^{-3}\mathbf{I})$, $g_c\sim\mathcal{CN}(0, 1)$, $g_e\sim\mathcal{CN}(0,5^{-3})$,
and $g_v\sim\mathcal{CN}(0,10^{-3})$. The transmit power constraint to noise power ratio is $P/\sigma^2=30$ dB. The target outage probability for signals from the BD to $U_c$ is $\epsilon=0.1$.

In Fig. 2, we present the convergence behavior of our proposed algorithm under different secrecy rate constraints at $U_c$ and $U_e$, where the reflection coefficient is $\alpha=0.5$. From Fig. 2, it is observed that our proposed algorithm converges after about 12 iterations.

\begin{figure}
\centering
\includegraphics[width=3.6in]{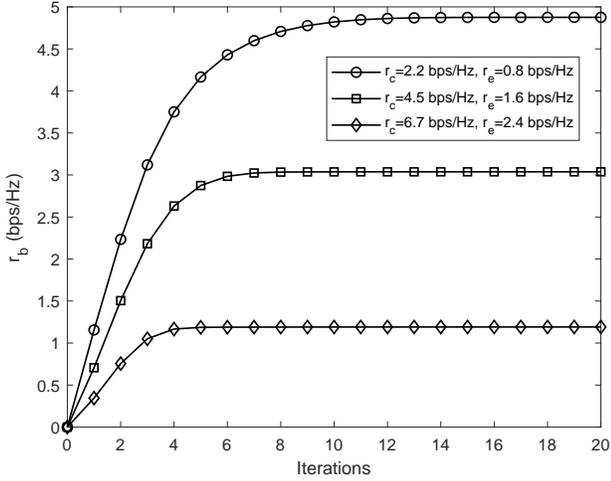}
\caption{$\epsilon$-outage secrecy rate $r_b$ versus the number of iterations; convergence performance comparisons under different achievable secrecy rate constraints $r_c$ and $r_e$.}
\end{figure}

In Fig. 3, we compare the secrecy rate region achieved by our proposed NOMA downlink BD aided SR network and the orthogonal multiple access (OMA) network. For the OMA network, we mean that the time-division multiple access (TDMA) scheme is employed and the signals transmitted to $U_c$ and $U_e$ are over two orthogonal time slots. From Fig. 3, it is observed that our proposed network achieves a much larger secrecy rate region than the OMA network.

\begin{figure}
\centering
\includegraphics[width=3.6in]{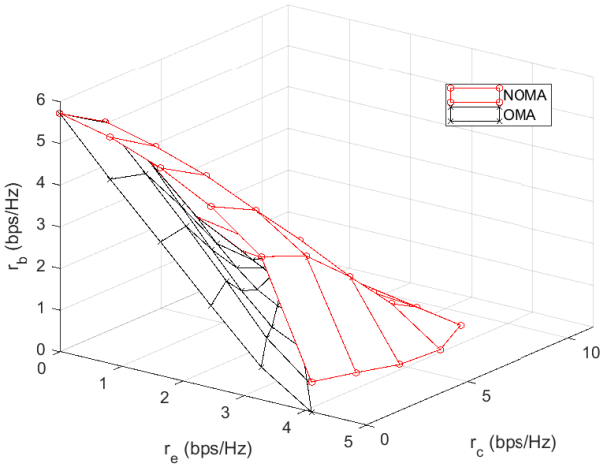}
\caption{Secrecy rate region comparison of our proposed NOMA downlink BD aided SR network and the OMA network.}
\end{figure}

In Fig. 4, we present the $\epsilon$-outage secrecy rate from the BD to $U_c$ for different reflection coefficients $\alpha$. From Fig. 4, it is found that when $\alpha$ is small, with the increase of $\alpha$, the $\epsilon$-outage secrecy rate increases. When $\alpha$ exceeds a certain value, the $\epsilon$-outage secrecy rate may increase or decrease with the increase of $\alpha$. This is because larger $\alpha$ means large interferences for signals from the BS to $U_c$ and $U_e$. If the signals from the BS to $U_c$ and $U_e$ cannot be successfully decoded, so do the signals from the BD to $U_c$.

\begin{figure}
	\centering
	\includegraphics[width=3.6in]{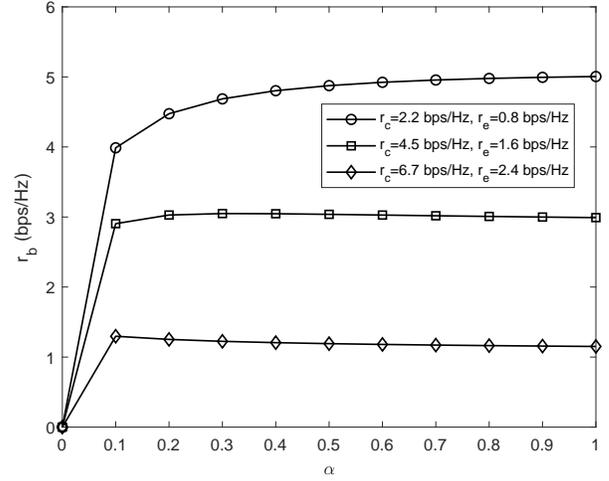}
	\caption{$\epsilon$-outage secrecy rate versus the reflection coefficient, $\alpha$; performance comparisons under different achievable secrecy rate constraints $r_c$ and $r_e$.}
\end{figure}

\section{Conclusion}
In this paper, we have proposed a CCCP-based $\epsilon$-outage secrecy rate optimization algorithm for the secure beamforming design in a downlink MISO NOMA BD aided SR network. Simulation results have shown that our proposed network achieves a much larger secrecy rate region than the OMA network.

\end{document}